\documentclass[sigconf,natbib=true]{acmart}

\usepackage[noend]{algorithmic}
\usepackage[ruled]{algorithm2e}
\usepackage{adjustbox}
\usepackage{enumitem}
\usepackage{multirow}

\copyrightyear{2024}
\acmYear{2024}
\setcopyright{rightsretained}
\acmConference[CIKM '24]{Proceedings of the 33rd ACM International Conference on Information and Knowledge Management}{October 21--25, 2024}{Boise, ID, USA}
\acmBooktitle{Proceedings of the 33rd ACM International Conference on Information and Knowledge Management (CIKM '24), October 21--25, 2024, Boise, ID, USA}
\acmDOI{10.1145/3627673.3679977}
\acmISBN{979-8-4007-0436-9/24/10}

\makeatletter
\gdef\@copyrightpermission{
  \begin{minipage}{0.3\columnwidth}
   \href{https://creativecommons.org/licenses/by/4.0/}{\includegraphics[width=0.90\textwidth]{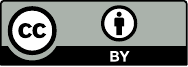}}
  \end{minipage}\hfill
  \begin{minipage}{0.7\columnwidth}
   \href{https://creativecommons.org/licenses/by/4.0/}{This work is licensed under a Creative Commons Attribution International 4.0 License.}
  \end{minipage}
  \vspace{5pt}
}
\makeatother

\DeclareMathOperator*{\argmax}{arg\,max}

\AtBeginDocument{%
  \providecommand\BibTeX{{%
    \normalfont B\kern-0.5em{\scshape i\kern-0.25em b}\kern-0.8em\TeX}}}

\newcommand{\msmarco}{\textsc{Ms Marco}\xspace}
\newcommand{\nq}{\textsc{NQ}\xspace}
\newcommand{\beir}{\textsc{Beir}\xspace}

\newcommand{\seismic}{\textsc{Seismic}\xspace}
\newcommand{\seismicpp}{\textsc{SeismicWave}\xspace}

\newcommand{\pisa}{\textsc{Pisa}\xspace}

\newcommand{\splade}{\textsc{Splade}\xspace}

\newcommand{\spladeT}{\textsc{Splade}-v3\xspace}

\newcommand{\cut}{\textsf{cut}}
\newcommand{\heapfactor}{\textsf{heap\_factor}\xspace}
\newcommand{\heap}{\textsc{Heap}}

\newcommand{\etal}{\emph{et al.}\xspace}

\newcommand{\knn}{$\kappa$-NN\xspace}

\begin{document}

\title{Pairing Clustered Inverted Indexes with \knn Graphs for Fast Approximate Retrieval over Learned Sparse Representations}

\author{Sebastian Bruch}
\affiliation{%
    \institution{Pinecone}
    \city{New York}
    \country{USA}
}
\email{sbruch@acm.org}

\author{Franco Maria Nardini}
\affiliation{%
    \institution{ISTI-CNR}
    \city{Pisa}
    \country{Italy}
}
\email{francomaria.nardini@isti.cnr.it}

\author{Cosimo Rulli}
\affiliation{%
    \institution{ISTI-CNR}
    \city{Pisa}
    \country{Italy}
}
\email{cosimo.rulli@isti.cnr.it}

\author{Rossano Venturini}
\affiliation{%
    \institution{University of Pisa}
    \city{Pisa}
    \country{Italy}
}
\email{rossano.venturini@unipi.it}



\begin{abstract}
Learned sparse representations form an effective and interpretable class of embeddings for text retrieval. While exact top-$k$ retrieval over such embeddings faces efficiency challenges, a recent algorithm called \seismic has enabled remarkably fast, highly-accurate approximate retrieval. \seismic statically prunes inverted lists, organizes each list into geometrically-cohesive blocks, and augments each block with a summary vector. At query time, each inverted list associated with a query term is traversed one block at a time in an arbitrary order, with the inner product between the query and summaries determining if a block must be evaluated. When a block is deemed promising, its documents are fully evaluated with a forward index. \seismic is one to two orders of magnitude faster than state-of-the-art inverted index-based solutions and significantly outperforms the winning graph-based submissions to the BigANN 2023 Challenge. In this work, we speed up \seismic further by introducing two innovations to its query processing subroutine. First, we traverse blocks in order of importance, rather than arbitrarily. Second, we take the list of documents retrieved by \seismic and ``expand'' it to include the neighbors of each document using an offline $\kappa$-regular nearest neighbor graph; the expanded list is then ranked to produce the final top-$k$ set. Experiments on two public datasets show that our extension, named \seismicpp, can reach almost-exact accuracy levels and is up to $2.2\times$ faster than \seismic.
\end{abstract}

\ccsdesc[500]{Information systems~Retrieval models and ranking}

\keywords{Learned sparse representations, maximum inner product search, inverted index.}

\maketitle


\section{Introduction}
\label{sec:introduction}
Learned sparse retrieval (LSR)~\cite{epic,splade-sigir2021,formal2021splade,formal2022splade,lassance2022efficient-splade} is a family of widely-used techniques that encode an input into \emph{sparse} embeddings---a vector whose dimensions correspond with terms in some dictionary, with nonzero coordinates indicating that the corresponding terms are semantically relevant to the input. Similarity is typically determined by inner product, so that
retrieval becomes the problem known as Maximum Inner Product Search (MIPS)~\cite{bruch2024foundations}: Finding the top-$k$ vectors whose inner product with a query vector is maximal.

Several reasons motivate research on LSR. First, its effectiveness is often on par with \emph{dense retrieval}---which learns dense embeddings~\cite{DBLP:series/synthesis/2021LinNY,karpukhin-etal-2020-dense,xiong2021approximate,reimers-2019-sentence-bert,santhanam-etal-2022-colbertv2,colbert2020khattab,10.1007/978-3-031-56060-6_1}. Importantly, studies show that LSR generalizes better to out-of-distribution collections~\cite{bruch2023fusion,lassance2022efficient-splade}. Second, sparse embeddings inherit many benefits of classical lexical models such as BM25~\cite{bm25original} and are amenable to well-understood algorithms and data structures such as the inverted index. Finally, because each dimension maps to a term, sparse embeddings are \emph{interpretable}.

Despite their attractive properties, MIPS over sparse embeddings faced significant efficiency
challenges~\cite{bruch2023sinnamon,bruch2023bridging,mackenzie2021wacky,mallia2022guided-traversal}. Recognizing this handicap, the $2023$ BigANN challenge at NeurIPS hosted a sparse retrieval track, which evaluated the accuracy-throughput trade-off of submitted solutions on the \splade~\cite{formal2023tois-splade} embeddings of \msmarco~\cite{nguyen2016msmarco}. Results were surprising: the winners were not inverted index-based algorithms, but an adaptation of HNSW~\cite{hnsw2020}, a graph-based approximate nearest neighbor (ANN) algorithm.

Motivated by BigANN, Bruch \etal proposed \seismic~\cite{bruch2024seismic}, an approximate sparse MIPS algorithm that is highly-accurate yet remarkably fast. In contrast to winning entries of BigANN, \seismic operates on two classic data structures: the inverted and the forward index. The crucial innovation in \seismic is that, each of its inverted lists is organized into geometrically-cohesive blocks, and each block is equipped with a \emph{summary} of the vectors contained in it. Note that, blocks are arranged arbitrarily within each list.

\seismic executes a term-at-a-time strategy to produce the top-$k$ approximate set for a query $q$. When consuming an inverted list, \seismic first computes the inner product between $q$ and every summary in that list to produce a ``potential'' score for each block. It then visits blocks in arbitrary order, and compares their potential with the smallest score in the top-$k$ heap to determine if a block should be further evaluated. If a block's potential exceeds the threshold, \seismic resorts to the forward index to compute the exact inner product between $q$ and every document in that block. Because summaries allow the query processor to skip over a large number of blocks, the algorithm saves substantial computation.

In this paper, we enhance \seismic by incorporating two ideas into the query processor to reach higher per-query latency, resulting in a method we name \seismicpp. Our first contribution is to visit blocks in order of their potential, as opposed to the arbitrary order in which they happen to appear in an inverted list.

Second, we leverage the \emph{clustering hypothesis} (CH)~\cite{JARDINE1971217}, which suggests that closely-related documents tend to be relevant to the same queries. We do so by introducing a $\kappa$-regular nearest neighbor graph (i.e., a directed graph where each document is connected to its $\kappa$ nearest neighbors) to ``expand'' the list of documents returned by \seismic. Specifically, we obtain the neighbors of each retrieved document to form an expanded set, and rank the resulting set to extract the top-$k$ subset. This pre-computed structure augments the inverted index---whose lists and blocks capture local similarities---with global information about the similarity between documents.

We must note that, a number of works have paired the inverted index with some realization of CH. MacAvaney \etal~\cite{10.1145/3511808.3557231}, for example, coupled a standard re-ranking step with a graph-based adaptive exploration, to add to the pool documents that are most similar to the highest-scoring documents. In Lexically-Accelerated Dense Retrieval (LADR), Kulkarni \etal~\cite{10.1145/3539618.3591715} use lexical retrieval techniques to seed dense retrieval with a document proximity graph. Our work is related to the above as we also use CH to speed up sparse MIPS.


\section{Methodology}
\label{sec:methodology}

In this section, we describe our contributions in detail. To make the discussion more concrete, we quickly describe our notation. We denote vectors with lower-case letters (e.g., $u$ and $v$) and reserve $q$ for the query. We use subscripts to denote specific coordinates (e.g., $q_i$ is the $i$-th coordinate of $q$). We write $\textit{nz}(u) = \{ i \;|\; u_i \neq 0 \}$ for the set of non-zero coordinates of a vector.

\begin{figure}[t]
\centering
\includegraphics[width=0.6\columnwidth]{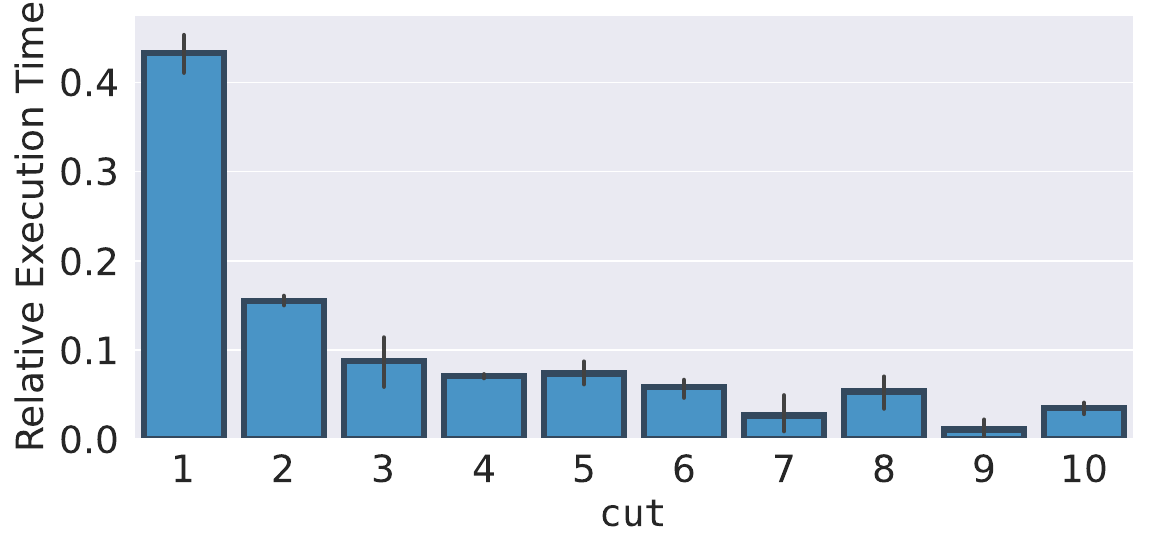}
	\caption{Breakdown of \seismic's query processing time per inverted list on \splade embeddings of \msmarco.}
	\label{fig:per_list_bkdown}
 \vspace{-4mm}
\end{figure}

\subsection{Overview of Query Processing in \seismic}
We briefly review \seismic in this section. However, as our review is concise due to space constraints, we refer the interested reader to~\cite{bruch2024seismic} for a detailed description of the algorithm.

\seismic's index comprises of two data structures: a forward index and an inverted index. The forward index is a mapping from document identifiers to raw vectors, and is used to compute exact inner products. The inverted index is made up of one inverted list per coordinate. Each inverted list keeps the id of the ``top'' $\lambda$ documents: order in the $i$-th list is determined by the value of the $i$-th coordinate. The $\lambda$ entries in an inverted list are then organized into $\beta$ geometric blocks by the application of a clustering algorithm to the document vectors in that list. Finally, each block is accompanied by a summary vector that represents its documents.

\seismic adopts a term-at-a-time query processing strategy. That means, the retrieval algorithm traverses the top \texttt{cut} inverted lists---where \texttt{cut} is a hyper-parameter---from $\textit{nz}(q)$. As it traverses each list, it inserts promising documents into a heap to maintain a current set of top-$k$ documents. Processing an inverted list involves visiting its $\beta$ blocks sequentially (in arbitrary order), computing the inner product of the block's summary vector with $q$, comparing the resulting score with the heap's threshold (scaled by \texttt{heap\_factor}, another hyper-parameter), and evaluating the documents within that block when the block's score exceeds the threshold. If a document's true score---computed using the forward index---exceeds the heap's current threshold, it is inserted into the heap.

\begin{figure}[t]
\centering
\includegraphics[width=0.6\columnwidth]{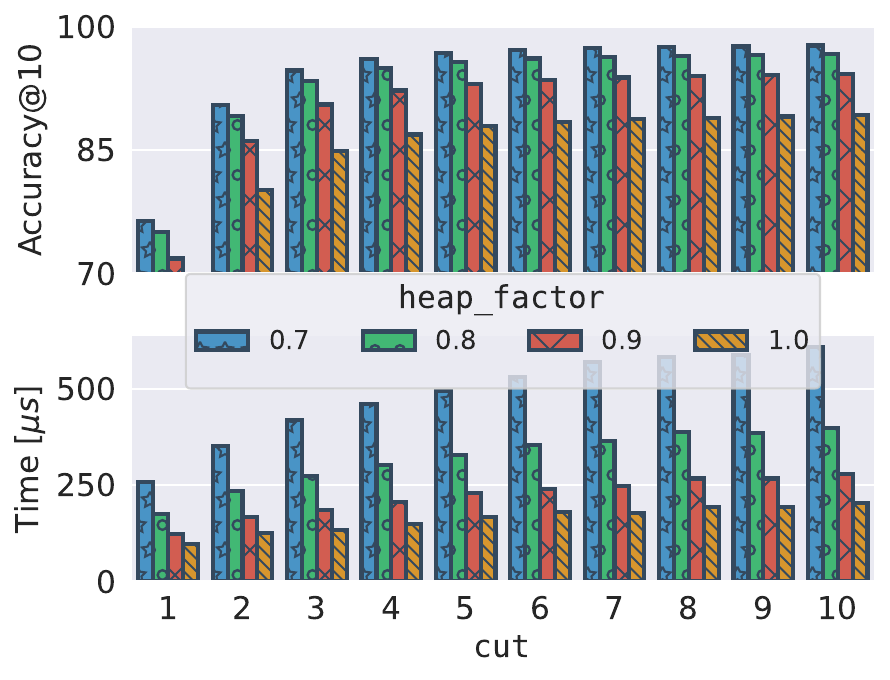}
	\caption{Top-$10$ accuracy of \seismic and search time versus the number of processed inverted lists on \splade embeddings of \msmarco. Accuracy rises quickly and plateaus, while latency increases steadily as more lists are processed.}
	\label{fig:slow_growth}
\vspace{-4mm}
\end{figure}

\subsection{Ordered Block Traversal: First Contribution}
The success of \seismic rests on the fact that the vast majority of blocks can be skipped without any processing beyond the inner product computation between their summary vectors and $q$. That is because the heap's threshold can only monotonically increase, so that, as more blocks are visited and evaluated, the likelihood that yet-unseen blocks qualify for further processing decreases. This phenomenon plays out in practice as shown in Figure~\ref{fig:per_list_bkdown}, which visualizes the proportion of \seismic's query processing time spent on the first $10$ inverted lists: More than $40\%$ of the search time is spent processing the first list, and about $60\%$ on the first two lists.

Our first contribution is to leverage this empirical property. In particular, rather than visiting blocks within the first inverted list in arbitrary order, we sort its blocks by their summary score (i.e., inner product of their summary vector with $q$) in descending order. The rest of the procedure remains the same. This change ensures that the most promising blocks are evaluated first, and the less promising blocks are more likely to be skipped.

Note that, inner products of $q$ with summaries in an inverted list can be computed efficiently with sparse matrix-vector multiplication. Sorting the blocks is also cheap as $\beta$ is typically small.

\subsection{\knn Graph: Second Contribution}
We make a second observation: As search for $q$ goes on, \seismic struggles to find good candidate blocks to evaluate. This is evident in Figure~\ref{fig:slow_growth}, which reports latency (in $\mu s$) and accuracy for $k=10$ as a function of the number of evaluated lists (\cut) for various values of \heapfactor. As is clear from the figure, \seismic's top-$k$ quality rapidly enters the high accuracy region ($>95\%$), but plateaus and struggles to reach almost-exact accuracy ($>98\%$). Naturally, exploring new inverted lists increases the execution time.

Our second contribution is to counter the phenomenon above by complementing the \seismic index with auxiliary information that allows it to \textit{refine} the candidate pool. In particular, we construct a \knn graph from a collection of documents, wherein each node represents a document and has an outgoing edge to $\kappa$ documents whose inner product with the source node is maximal. The resulting structure can be stored as a look-up table consisting of pairs of a document id and a list of its $\kappa$ closest neighbors. The \knn graph allows us to identify the nearest neighbors of a document quickly, realizing CH. Formally, let us denote by $\mathcal{N}(u)$ the set of $\kappa$ closest documents to document $u \in \mathcal{X}$, where $\mathcal{X}$ is the collection:
\begin{equation}
    \mathcal{N}(u) = \argmax^{(\kappa)}_{v \in \mathcal{X}} \; \langle u, v \rangle.
    \label{equation:mips}
 \end{equation}

We use the \knn graph as follows. Once \seismic concludes its search for the top-$k$ documents, we take the set of documents in the heap and denote it by $\mathcal{S}$. We then form the \emph{expanded set} $\tilde{\mathcal{S}} = \bigcup_{u \in \mathcal{S}} \big( \{ u \} \cup \mathcal{N}(u) \big)$, compute scores for documents in $\tilde{\mathcal{S}}$, and return the top-$k$ subset. This procedure is shown in Algorithm~\ref{algorithm:retrieval}.

\begin{algorithm}[t]
\SetAlgoLined
{\bf Input: }{$q$: query;
$k$: number of results;
$\mathcal{N}(\cdot)$: \knn graph represented as a function that returns $\kappa$ neighbors of its argument;
\heap: heap with the top-$k$ results from \seismic.}\\
\KwResult{ A \heap\ with the top-$k$ documents.}
\begin{algorithmic}[1]
\STATE $S \leftarrow$ the ids of the vectors in the \heap
\FOR{$u \in S$}
    \FOR{$v \in \mathcal{N}(u)$}
        \STATE $p = \langle q, {\sf ForwardIndex}[v] \rangle$
        \IF{$\heap.{\sf len()} < k$ or $p > \heap.{\sf min()}$ }
            \STATE \heap.{\sf insert}$(p, v)$
        \ENDIF
        \IF{$\heap.{\sf len()} = k+1$}
            \STATE \heap.{\sf pop\_min()}
        \ENDIF
    \ENDFOR
\ENDFOR
\RETURN \heap
\end{algorithmic}
\caption{Refining \seismic's results with \knn graph.\label{algorithm:retrieval}}
\end{algorithm}


\section{Experiments}
\label{sec:experiments}

\noindent \textbf{Datasets}.
We experiment on two publicly-available datasets: \msmarco{} v1
Passage~\cite{nguyen2016msmarco} and Natural Questions (\nq{}) from \beir~\cite{thakur2021beir}.
\msmarco{} is a collection of $8.8$M passages in English. In our evaluation,
we use the smaller ``dev'' set of $6{,}980$ queries.
\nq{} is a collection of $2.68$M questions in English and a ``test'' set of $3{,}452$ queries.

\vspace{1mm}
\noindent \textbf{Learned Sparse Representations}.
We evaluate all methods on embeddings generated by two LSR models:
\begin{itemize}[leftmargin=*]
\item \splade~\cite{formal2022splade}. Each non-zero entry is the importance weight of a term in the BERT~\cite{devlin2019bert}
WordPiece~\cite{wordpiece} vocabulary consisting of $30$,$000$ terms.
We use the \texttt{cocondenser-ensembledistil}
version of \splade that yields MRR@10 of $38$.$3$ on the \msmarco dev set.
The number of non-zero entries in documents (queries) is, on average,
$119$ ($43$) for \msmarco and $153$ ($51$) for \nq.
\item \spladeT~\cite{lassance2024spladev3}. An improved variant of \splade that incorporates a modified objective and distillation strategy. It yields MRR@10 of $40{.}3$ on the \msmarco dev set. The number of non-zero entries in documents (queries) is, on average, $168$ ($24$) for \msmarco.
\end{itemize}

\vspace{1mm}
\noindent \textbf{\knn Graph}. Constructing the (exact) \knn graph is expensive due to its quadratic time complexity. As such, we relax the construction to an approximate (but almost-exact) \knn graph: each document is connected to its approximate set of top-$\kappa$ documents. To that end, we use a \seismic index with the following parameters to retrieve the top-$\kappa$ candidates for every document in the collection: $\lambda = 10{,}000$, $\beta=2{,}000$, $\alpha=0.6$, $\cut=15$, $\heapfactor=0.7$. Note that, the \knn graph is formed only once for a collection. Storing the \knn graph takes 
$ (\lfloor \log_{2} (n-1)\rfloor  + 1)n\kappa$ bits, where $n$ is the size of the collection.

\vspace{1mm}
\noindent \textbf{Baselines}. The original work by Bruch \etal~\cite{bruch2024seismic} reports a wide gap between \seismic and all other state-of-the-art sparse retrieval algorithms. As such, rather than comparing our extension of \seismic with all baselines, we only contrast \seismicpp with \seismic.

\vspace{1mm}
\noindent \textbf{Hyperparameters}. 
When building \seismic and \seismicpp indexes, we first fix a memory budget as a multiple of the size of the forward index. We then sweep the hyper-parameters as follows to find the best configuration that results in an index no larger than the budget: $\lambda \in \{2000, 2500, 3000, 4000, 5000, 6000\}$, $\beta \in \{ \lambda/10, \lambda/5 \}$, $\alpha \in \{ 0.4, 0,5, 0.6 \}$, and for \seismicpp, $\kappa \in \{ 10, 20, 30, 40, 50\}$. We set $\cut \in \{1,2,3,4,5,6,7,8,10, 12,14 \}$ and $ \heapfactor \in \{0.7, 0.8, 0.9, 1.0 \}$, and report the best configuration.

\vspace{1mm}
\noindent \textbf{Metrics}. We use three metrics to evaluate all methods:
\begin{itemize}[leftmargin=*]
    \item Latency ($\mu$sec.). The time to retrieve top-$k$ vectors given a query in single-thread mode.
    Latency does not include embedding time.
    \item Accuracy. Percentage of true nearest neighbors recalled.
    \item Index size (MiB). The space the index occupies in memory.
\end{itemize}

\noindent \textbf{Hardware Details}.
We implemented the methods in Rust and compile using Rust version $1{.}77$ with the highest level of optimization made available by the compiler. We conduct experiments on a server equipped with one Intel i9-9900K CPU with a clock rate of $3{.}60$ GHz and $64$ GiB of RAM. The CPU has $8$ physical cores and $8$ hyper-threaded ones. We query the index using a single thread.

\begin{table*}[t]
	\centering
    \adjustbox{max width=\textwidth}{%
   	\begin{tabular}{cccllllllllll}
        \toprule
         Embeddings & Budget & Accuracy (\%) & 90 & 91 & 92 & 93 & 94 & 95 & 96 & 97 & 98 & 99 \\
        \midrule
        \multirow{4}{*}{\splade} &

        \multirow{2}{*}{$1.5 \times$} &
        
        \seismic
        & $179$ ($1.5\times$) & $195$ ($1.7\times$) & $234$ ($1.9\times$) & $277$ ($2.0\times$) & $293$ ($2.0\times$) & $370$ ($1.9\times$) & $531$ ($2.2\times$) & $-$ & $-$ & $-$ \\

        & & \textbf{\seismicpp} & 116 & 116 & 121 & 141 & 148 & 195 & 237 & -- & -- & -- \\
        
        \cmidrule{2-13}
        
        &\multirow{2}{*}{$2 \times$}  & \seismic
        & $206$ ($1.3\times$) & $206$ ($1.3\times$) & $206$ ($1.3\times$) & $206$ ($1.3\times$) & $233$ ($1.4\times$) & $257$ ($1.4\times$) & $305$ ($1.6\times$) & $336$ ($1.6\times$) & $441$ ($1.7\times$) & $669$ ($1.6\times$) \\
        & & \textbf{\seismicpp} & $ 160 $ & $ 160 $ & $ 160 $ & $ 160 $ & $160  $ & $177  $ & $193$ & $218$ & $258$ & $409$ \\

        \midrule
        \multirow{4}{*}{\spladeT} &
        \multirow{2}{*}{$1.5 \times$} &
        \seismic &
        $183 $ ($1.3\times$) & $205 $ ($1.3\times$) & $223 $ ($1.4\times$) & $261 $ ($1.6\times$) & $ 290$ ($1.8\times$) & $323$ ($1.7\times$) & $417 $ ($1.7\times$) & $ 612$ ($2.1\times$) & -- & -- \\

        & & \textbf{\seismicpp} & $ 131 $ & $ 159 $ & $ 159 $ & $ 159 $ & $ 159 $ & $ 186 $ & $ 234 $ & $ 296 $ & $470$ & -- \\
        \cmidrule{2-13}
        & \multirow{2}{*}{$2 \times$}&
        \seismic &
        $213 $ ($.93\times$) & $261 $ ($1.1\times$) & $261 $ ($1.1\times$) & $261 $ ($1.1\times$) & $ 261$ ($1.1\times$) & $ 319 $ ($1.4\times$) & $319 $ ($1.4\times$) & $ 425$ ($1.6\times$) & $ 519 $ ($1.9\times$) & $ 806 $ ($1.7\times$)
        \\
        
        & & \textbf{\seismicpp} & $ 229 $ & $ 229 $ & $ 229 $ & $ 229 $ & $ 229 $ & $ 229 $ & $ 229 $ & $ 270 $ & $270 $ & $ 460$ \\

        \bottomrule
    \end{tabular}}
	\caption{Mean latency ($\mu$sec/query) at different accuracy cutoffs with speedup (in parenthesis) gained by \seismicpp on \msmarco. The ``Budget'' column indicates the memory budget as a multiple of the size of the forward index.\label{table:results1}}
    \vspace{-5mm}
\end{table*}

\subsection{Results}
\begin{figure}[t]
\centering
\includegraphics[width=0.9\columnwidth]{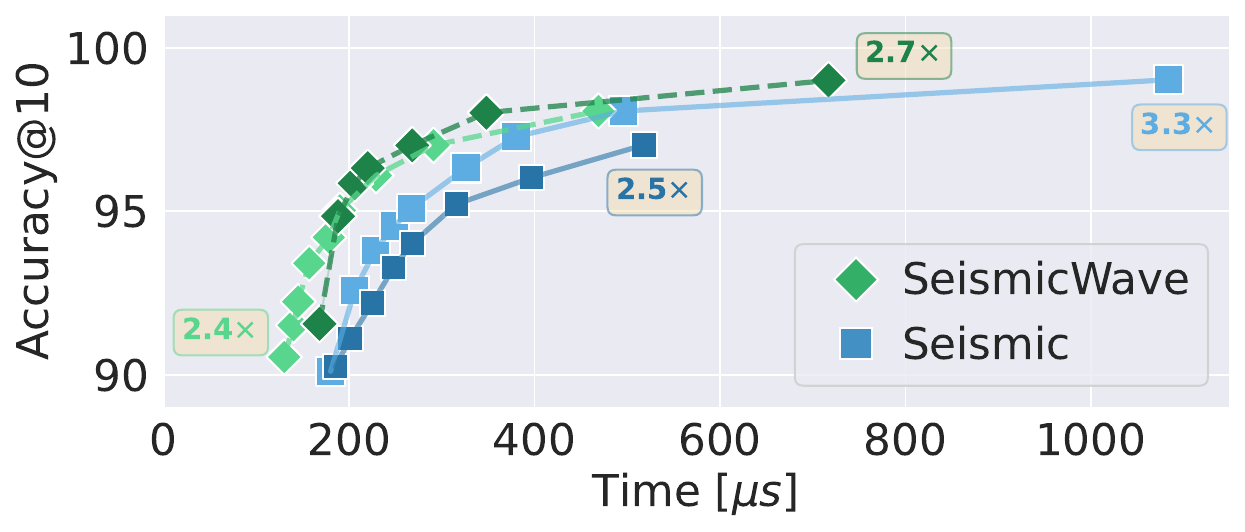}
	\caption{Comparison of \seismic and \seismicpp by memory, latency, and accuracy on \splade embeddings of \nq.}
	\label{fig:option2}
 \vspace{-3mm}
\end{figure}

\noindent \textbf{\msmarco}. Table~\ref{table:results1} compares \seismicpp and \seismic on \msmarco. For each embedding type, we consider two memory budgets expressed as multiples of the size of the collection ($4$GB for \splade and $5{.}6$GB for \spladeT). We choose the best index configuration that respects the budget and report the fastest configuration reaching the accuracy cutoffs from $90\%$ to $99\%$. 
We also report the speedup in parenthesis gained by \seismicpp over \seismic.

\seismicpp outperforms \seismic in all the evaluation scenarios. On \splade, \seismicpp achieves $2.2\times$ speedup in the $1.5\times$ setting, and is up to $1.7\times$ faster than \seismic in the $2.0\times$ scenario.
On \spladeT, \seismicpp outperforms \seismic in both memory budgets. As \spladeT embeddings are less sparse, the memory impact of storing the \knn graph is smaller. 

Interestingly, the ``exact'' algorithm \pisa~\cite{MSMS2019} caps at $99\%$ accuracy due to quantization, which enables significant memory savings. At $99\%$ cutoff, \seismicpp takes $409 \mu s$; $1.6\times$ faster than $\seismic$ and \emph{two orders of magnitude} faster than \pisa, which takes $95{,}818 \mu s$.

\smallskip 
\noindent \textbf{\nq}. We also compare the two methods on \nq. The memory budgets used for \msmarco---$1.5\times$ and $2\times$---are insufficient to achieve satisfying performance with \seismic. We speculate this to be due to the distributional differences between the two datasets: \nq is about $3\times$ smaller but has more non-zero entries per embedding.

Figure~\ref{fig:option2} compares \seismic and \seismicpp by latency and accuracy with two indexes per solution. We annotate each solution with its memory budget as a multiple of the size of the forward index. \seismicpp is up to $1.8\times$ faster than \seismic with a slightly lower memory budget ($2.4\times$ vs $2.5\times$). Additionally, \seismicpp reaches $99\%$ with a budget of $2.7\times$ and a speedup of $1.5\times$, whereas \seismic requires a budget of $3.3\times$ to reach the same accuracy---a saving of approximately $30\%$ of the inverted index overhead.

\subsection{Ablation Study}
\begin{figure}[t]
\centering
\includegraphics[width=0.9\columnwidth]{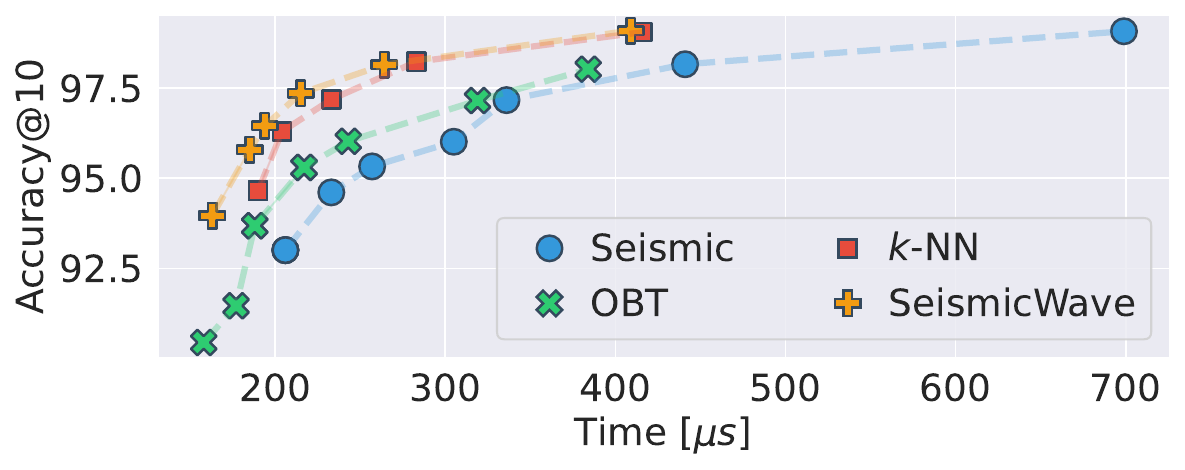}
	\caption{Impact of components of \seismicpp with memory budget of $2\times$ ($8$GB) on \splade embeddings of \msmarco.}
	\label{fig:ablation}
 \vspace{-1mm}
\end{figure}

In Figure~\ref{fig:ablation}, we break down the impact of our contributions on the \splade embeddings of \msmarco with a memory budget of $2\times$ (i.e., $8$GB). In this figure, OBT denotes \seismic with Ordered Block Traversal; \knn refers to \seismic with the \knn graph; and, \seismicpp is the combination of \seismic, OBT, and \knn graph.

It is clear that, OBT improves the efficiency of \seismic for lower accuracy cutoffs, but may slightly reduce the maximum achievable accuracy as a result of skipping more blocks in the first list. This degradation is well-compensated by the addition of the \knn graph, which reaches almost-exact search. The combination of OBT and the \knn graph (\seismicpp) yields the best performance overall, with a further improvement of $10\%$ over the \knn graph.

\section{Conclusions and Future Work}
\label{sec:conclusions}
We introduced two algorithmic improvements to \seismic~\cite{bruch2024seismic}, a state-of-the-art retrieval algorithm designed for learned sparse representations. In particular, we modified how inverted lists are traversed, and augmented \seismic with global information in the form of a \knn graph. As we empirically demonstrated, fo a fixed memory budget, our method, named \seismicpp, outperforms \seismic's latency by a factor of $2$ or less. The advantage of \seismicpp is especially evident if a higher accuracy is desired.

We did not use any form of compression in \seismicpp. We leave an empirical examination of the effect of compression and quantization on index size, latency, and accuracy to future work.

\vspace{1mm}
\section{Acknowledgements}
This work was partially supported by the Horizon Europe RIA ``Extreme Food Risk Analytics'' (EFRA), grant agreement n. 101093026, by the PNRR - M4C2 - Investimento 1.3, Partenariato Esteso PE00000013 - ``FAIR - Future Artificial Intelligence Research'' - Spoke 1 ``Human-centered AI'' funded by the European Commission under the NextGeneration EU program, by the PNRR ECS00000017 Tuscany Health Ecosystem Spoke 6 ``Precision medicine \& personalized healthcare'' funded by the European Commission under the NextGeneration EU program, by the PNRR IR0000013 ``SoBigData.it - Strengthening the Italian RI for Social Mining and Big Data Analytics'' funded by the European Commission under the NextGeneration EU program, by the MUR-PRIN 2017 ``Algorithms, Data Structures and Combinatorics for Machine Learning'', grant agreement n. 2017K7XPAN\_003, and by the MUR-PRIN 2022 ``Algorithmic Problems and Machine Learning'', grant agreement n. 20229BCXNW.

\newpage
\balance
\bibliographystyle{ACM-Reference-Format}
\bibliography{biblio}

\end{document}